\documentclass[]{article}

\usepackage[utf8]{inputenc}

\usepackage{fullpage}

\usepackage{amsmath}
\usepackage{amssymb}

\usepackage{boxedminipage}

\usepackage{listings}
\usepackage{minitoc}
\usepackage{ifpdf}
\pdfoutput=1

\ifpdf
\usepackage[pdftex]{graphicx}
\else
\usepackage{graphicx}
\fi
\title{
Is there enough fertile soil to feed a planet of growing cities?
}
\author{Roberto D'Autilia,\\
Formulas: Mathematics Laboratory,\\
Department of Architecture,\\
University of Roma3\\
roberto.dautilia@uniroma3.it\\
\\
Ilaria D'Ambrosi\\
Department of Architecture,\\
University of Roma3\\
ila.dambrosi@gmail.com}

\date{2014-16-3}

\begin{document}

\ifpdf
\DeclareGraphicsExtensions{.pdf, .jpg, .tif}
\else
\DeclareGraphicsExtensions{.eps, .jpg}
\fi

\maketitle

\begin{abstract}
We analyze a scaling law for the consumption of agricultural soil by cities.
The nonlinear dependence of the size of the city on the number of inhabitants gives rise to an equation for population dynamics.
We found the asymptotic limit of the solution for this equation, given by the carrying capacity in terms of number of inhabitants that can be fed.
The carrying capacity as a function of the scaling law exponent is computed numerically, showing that the exponent must be very small to ensure a food sustainability.
We suggest a bound for the value of this exponent and analyze the reliability of the scaling law for major cities.
\end{abstract}

\section{Introduction}
In recent papers \cite{citeulike:1248679} \cite{MicBat} \cite{Betten2013} the power law  $Y=Y_0N^\beta$ has been assumed to relate the city population $N$ with a general city development resource $Y$.
For the population dynamics driven by the resource $Y$, this dependence gives a Bernoulli equation with non-integer exponent:
\begin{equation}
	Y_0N(t)^\beta=R N(t)+E{dN\over dt}(t)
\label{bern}
\end{equation}
where $R>0$ is the fraction of $Y$ required to maintain an individual per unit time,  $E>0$ is the fraction needed to add a new one per time, and $Y_0$ is a normalization constant.
For $0<\beta<1$ the solution of the equation (\ref{bern})
leads to a  sustainable behavior (scale economy) characterized by a finite carrying capacity for the population $N(t)$, while for $\beta>1$ the demographic trend is superlinear \cite{citeulike:1248679}.
The value $\beta=1$ therefore represents the threshold between sustainability (scale economy) and non-sustainability (non-scale economy) for the urban growth driven by the resource $Y$.

Among the development quantities, the city area size is indicative of the relationship between sustainability and economical development. 
World colonization  started about 50,000 years ago \cite{citeulike:5947116} and was in fact progressively characterized by a rapid urbanization process corresponding to a loss of agricultural land.
In this framework we consider the development quantity ``urban land cover'' $Y=Y(N)$ given by the soil covered by buildings, roads and, in general, by all urban infrastructures.

The global area of cities presently covers almost 0.5\% of the planet's land area, and 3.93\% of the arable land \cite{citeulike:9394452}, but in 2008, for the first time in human history, more people were living in cities than in rural areas.
Although 3.93\% of arable land could be considered a small percentage, it does not imply that there is still a large amount of land for buildings, as food demand depends on the density of people living in cities. A denser city needs more food from the surrounding countryside than a larger but less populated one.

2008 not only was the year of global urbanization, it was also the year of a major food-price crisis.
The causes of the 2007-2008 food crisis are many and are still under analysis \cite{RePEc:fpr:resrep:165}, but the crisis itself suggests we address our attention to the relationship between urbanization policy and food availability for the next few years.

According to recent projections \cite{citeulike:9394452}, the world urban land cover in developing countries will increase from 300,000 $km^2$ in 2000 to 770,000 $km^2$ in 2030 and 1,200,000 $km^2$ in 2050.
In the perspective of a planet of cities \cite{citeulike:9394452}, it is natural to ask how much the cities can grow in terms of people and area, while preserving enough country to produce food for the whole world population.
The lack of food near cities can be cause of social conflicts: when local food is scarce, the inhabitants will be forced to import food from somewhere else, with the risk of regional conflicts.

As noticed in \cite{citeulike:9445879}, {\sl ``geographic proximity has been related to sustainability for a variety of reasons, encompassing the ecological, economic and social dimensions of the food system. For example, in local food systems where producers and consumers are in closer physical proximity, local food is presumed to travel shorter distances and consequently reduce the amount of energy used and greenhouse gas emissions released in the transport of foods.''}
The question to ask is then {\sl ``To what degree can food be produced locally?''} \cite{citeulike:9445879} and how much local food production will be constrained by the anthropization rate and the loss of agricultural soil.

In this paper we considered only rural areas as primary sources of food, excluding seafood, biofuel and other non-food agricultural products. 
The growth of urban areas decreases the availability of fertile soil, but the increment of city size corresponds to the increment of the population and to an increasing request for food-production land.
The equilibrium of this process depends on the law relating the population to the urban area.
The knowledge of this dependence allows one to determine the equilibrium of the city expansion process in terms of number of persons who can be fed by the surrounding country.
We assume that this dependence is given by a power law, a hypothesis that is confirmed, with a good approximation, by the experimental data \cite{10.1257/aer.101.5.2205}. 

We introduce a model to represent the possible evolution of small geographic areas containing cities, such as counties, as well as global agricultural soil and global city cover development.
In principle we could represent the world as a single agricultural area and consider the whole urban area of the planet as one big city.
However if we analyze large areas, the parameters of the model, depending on the fertility and on local diets, should be averaged on a large scale, whereas accurate comparison with experimental data can be better made on a small scale.

The availaibility of food is determined also by other quantities such as agricultural technologies and the availability of renewvable energy for transport.
We take into account this aspect by means of a parameter that correlates farmland with the number of people that can be fed by it.

The main result of our analysis is that for the ``urban land cover'' resource, the threshold $\beta=1$, suggested in \cite{citeulike:1248679} for generic resources, is a value too high to preserve sustainability, unless the value of $Y_0$ is very small.
We also propose a critical value $\bar\beta$ for the exponent of the power law, below which urban developement could be considered sustainable.

\section{The power law and population dynamics}
We call $C(t)$, with $\mathbb{R}_+\ni t\rightarrow C(t)\in\mathbb{R}_+$ the whole land available for both urbanization and food production, and assume the urban land cover to be a power law of the population $N$
\begin{equation}
\mathbb{R}_+\ni Y=Y_0N^\beta\in\mathbb{R}_+
\label{pow}	
\end{equation}
Putting $N=1$ this equation shows that the quantity $Y_0$ represents the urban soil needed for the first person and plays the role of a normalization constant.
For the urbanized soil, the law (\ref{pow}) shows how much of the city soil should be shared, and therefore also how much area is devoted to infrastructures.

Following eq. (\ref{pow}), the first inhabitant of the city needs $Y_0$ hectares of urban land, but the soil will be used for both housing and services.
In general a second inhabitant does not exploit $2Y_0$ hectares of soil, because s/he will use the same infrastructure as the first settler, and needs only $Y_0(2^\beta-1)$ hectares of new land, with $\beta<1$.

The difference $Y_0(2^\beta-1)$ is given mainly by the housing area for the second inhabitant with at most a possible small increment in services.
In general soil needed for housing grows faster than soil devoted to infrastructure.
For general urban development quantities, this analysis is confirmed by the data in \cite{citeulike:1248679}, where the value of $\beta$ was found to be about 1 for housing, 0.83 for road surface, 0.77 for the number of gasoline stations, and 0.87 for the length of electric cables.

If $C(t)$ is all arable land, the area available for food production is given by $C(t)-Y=C(t)-Y_0N^\beta$.
Following \cite{citeulike:1248679} at time $t$ a fraction $R(t)$ of this area is exploited to provide food for the population, and a fraction $E(t)$ to grow it per time.
The population dynamics (\ref{bern}) becomes:
\begin{equation}
	C(t)-Y_0N(t)^\beta=R(t) N(t)+E(t){dN\over dt}(t)
	\label{chini}
\end{equation}
which is a Chini-like equation \cite{chini1924} with $0<\beta<1$, where in general $C(t)$, $R(t)$ and $E(t)$ depend on time and possibly on other parameters.
In what follows we assume these (positive) quantities to be constant.
In section \ref{sviluppi} we also discuss some possible generalizations of equation (\ref{chini}).

The quantity $R$ in eq.(\ref{chini}) is the amount of soil needed to feed one person per time unit, and depends on eating behaviors and land fertility.
The annual diet has been measured for example for the New York State in terms of hectares needed to feed a person per year \cite{peters2007testing}.
We assume that the minimum amount of arable land to feed an adult is $0.18$ hectares: it corresponds to a diet with no meat and 65 grams of added fat per day.
The maximum amount of land used by a diet is 0.86 hectares per year and the corresponding diet has a daily intake of 381 grams of meat and 65 grams of added fat \cite{peters2007testing}.

Unlike the Bernoulli equation, the solution of the Chini equation can not be found in a general form for every $0<\beta<1$, but eq.(\ref{chini})  can be easily integrated numerically.
For constant coefficient $C$, $E$ and $R$, and in general when the Chini invariant \cite{kamke1956differentialgleichungen} is independent of $t$, the solution can be found for given $\beta$ in terms of implicit functions.

For example if $\beta=1/2$ the solution can be given as an inverse function of
\begin{equation}
	{2Y_0\arctan\big({Y_0+2R\sqrt{k}\over\sqrt{-4CR-Y_0^2}}\big)\over R\sqrt{-4CR-Y_0^2}}-{\ln(C-Y_0\sqrt{k}-Rk)\over R}
\end{equation}
where
\begin{equation}
	k={t\over E}+{2Y_0\arctan\big({2\sqrt{N(0)}R+Y_0\over\sqrt{-4CR-Y_0^2}}\big)-\sqrt{-4CR-Y_0^2}\ln\big(C-N(0)R-\sqrt{N(0)}Y_0\big)\over R\sqrt{-4CR-Y_0^2}}
\end{equation}
and $N(0)$ is the initial population.

\begin{figure}[ht]
\centering
\includegraphics[width=0.8\textwidth]{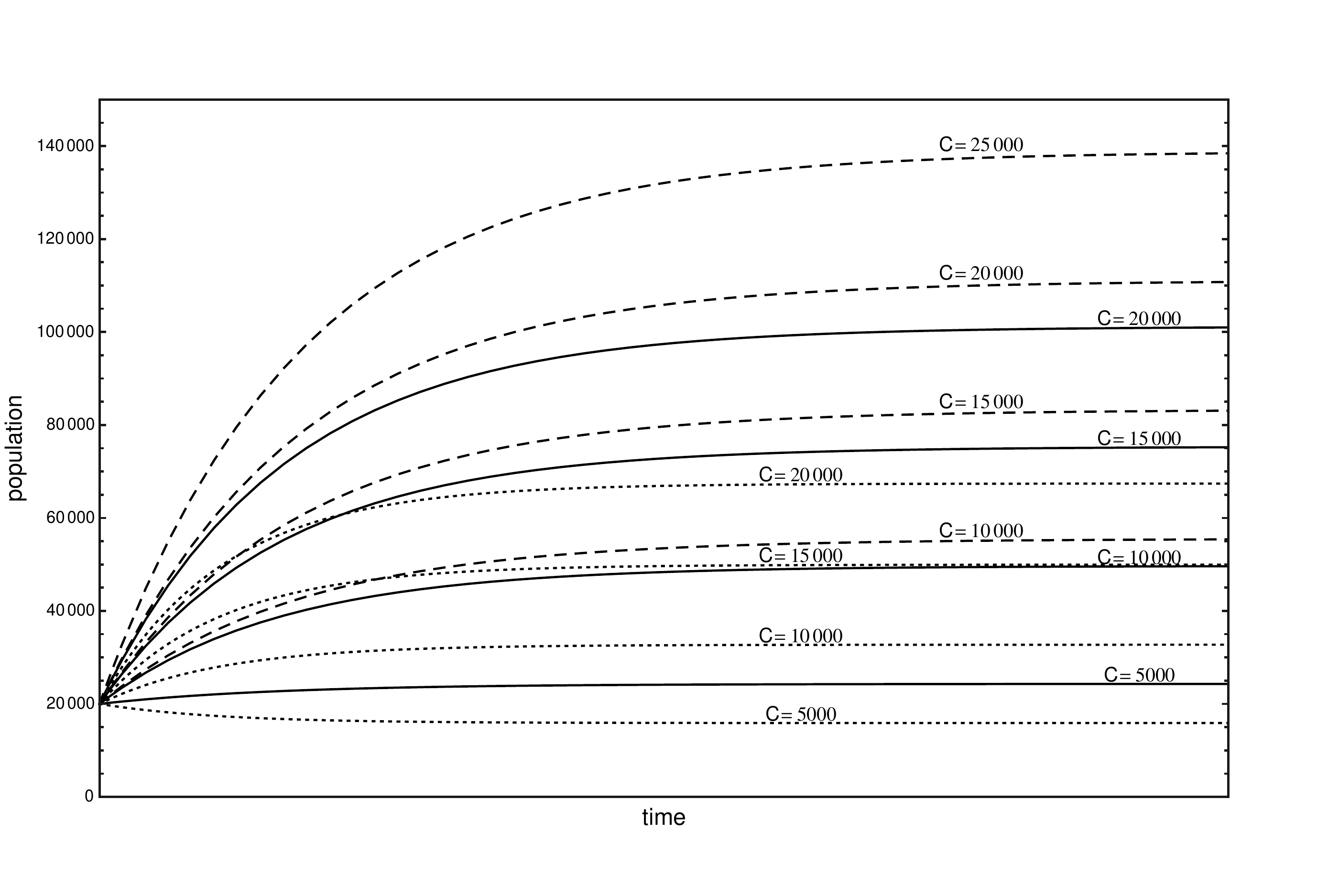}
\caption{The number of people that can be fed in an area $C$ as function of time for $\beta=0.3$ (dashed line) $\beta=0.74$ (continuous line)) and $\beta=0.9$ (dotted line)}
\label{fig:chini}
\end{figure}

The carrying capacity is given by the horizontal asymptote (if it exists) of the solution, representing the maximum number of people who can be fed:
\begin{equation}
	N_{eq}(\beta,Y_0)\equiv lim_{t\rightarrow\infty}N(t)	
\end{equation}
where we evidenced the explicit dependence on the two parameters $(Y_0,\beta)$.
Its value does not depend on the quantity $E$, which determines only how fast the asymptote is reached. 
The carrying capacity $N_{eq}(Y_0,\beta)$ is always bounded by the values $C/R$, corresponding to the limit situation of all land used for food production and no city in $C$.
Fig. \ref{fig:chini} shows the solution of (\ref{chini}) as a function of time, where we put $N(0)=20000$ as the initial population, the available soil ranging from 5000 to 25000 hectares,  $\beta=0.3, 0.74, 0.9$ and $Y_0=0.35$ hectares, corresponding to the value measured in northern Italy, similar to the values found for Japan.

Equation (\ref{chini}) is different from the logistic models like the Malthus-Vehrulst models \cite{murray2002mathematical}, as the carrying capacity depends on the parameter $\beta$ and is not given in the equation.
The exponent of the nonlinear term is the criticity parameter of the model.

\begin{figure}[ht]
\centering
\includegraphics[width=0.8\textwidth]{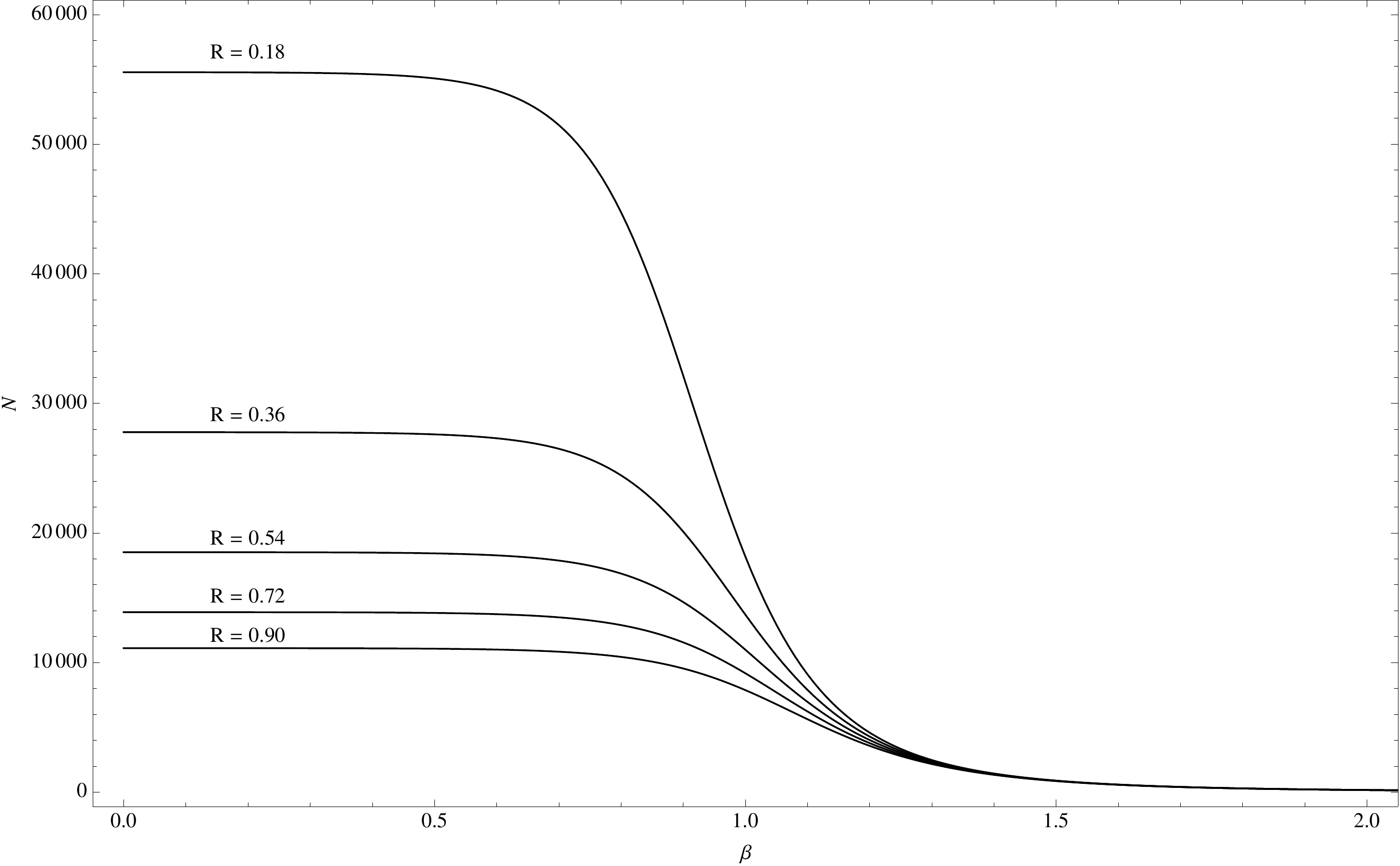}
\caption{The carrying capacity $N$ (number of people) as a function of $\beta$ for different diets $R=0.18, 0.36, 0.54, 0.72, 0.90$, $C=10000$ ha. and $Y_0=0.35$ ha.}
\label{fig:carcap}
\end{figure}

The asymptote $N_{eq}(\beta,Y_0)$ of the solution of (\ref{chini}) shows that, given $Y_0$, if $\beta$ is small enough, the same area can support a larger population.
Fig. (\ref{fig:carcap}) represents the carrying capacity $N_{eq}(\beta)$, with $Y_0=0.35$, for different values of the diet $R$ (in hectares) computed numerically from eq.(\ref{chini}).
When a city wants to plan a locavore food policy, to reduce for example the costs and the pollution derived from the hydrocarbons necessary for the food transportation, one must have not only an
adequate amount of agricultural soil around the city, but also an urban growth policy such that $\beta$ lies in the quasi linear regime of $N_{eq}(\beta)$.

Although we do not have an explicit form for $N_{eq}(\beta)$, the numerical results show that the carrying capacity is not linear in $\beta$ for any $Y_0$.
The shape of $N_{eq}(\beta,Y_0)$ shows that the use of the soil resource, when $\beta$ is small, does not significantly changes the carrying capacity.
On the contrary, when $\beta$ is bigger than a given value $\bar\beta$ that separates the linear from the quasi linear behavior (${\partial\over\beta} N_{eq}(\beta,Y_0)\approx0$), an urban policy that increase $\beta$ even a little, can lead to disastrous effects.

The nonlinear behavior of $N_{eq}(\beta)$ also has remarkable consequences for urban and rural planning.
If $\beta$ is smaller than the critical value $\bar\beta$ below which  $N_{eq}(\beta)$ is almost constant for a given $Y_0$, then the growth of the population does not influence too much the depletion of the agricultural resources.
However, if $\beta>\bar\beta$, the population that can be supported by the remaining agricultural land collapses rapidly to a very low value, forcing people to seek food elsewhere or the population to decline.

\begin{figure}[ht]
\centering
\includegraphics[width=0.8\textwidth]{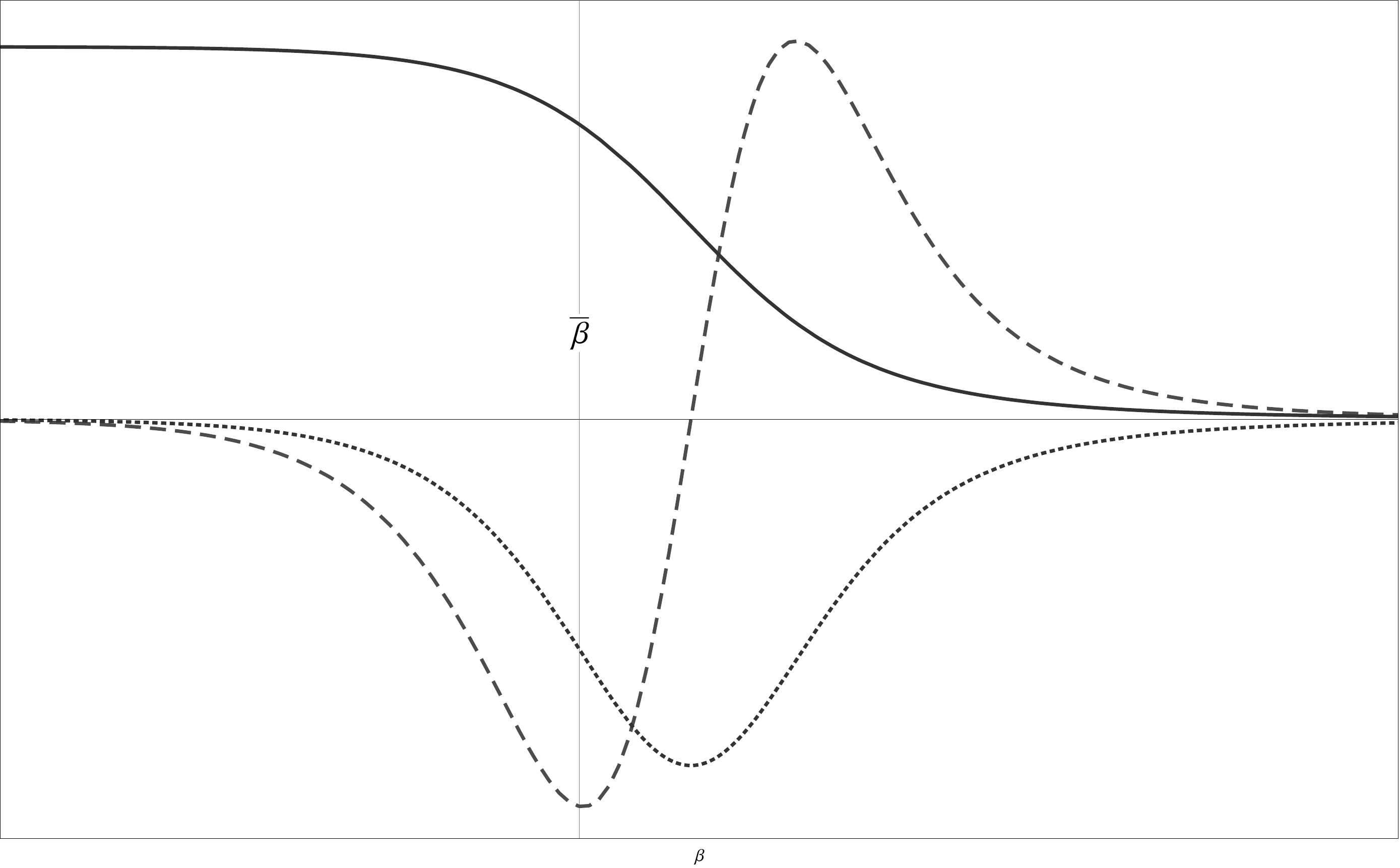}
\caption{The carrying capacity as a function of $\beta$ (continuous line) together with its derivative (dotted line) and second derivative (dashed line) in scaled units}
\label{fig:derivative}
\end{figure}

The $N_{eq}(\beta)$ function also depends on the parameter $R$.
Fig. \ref{fig:carcap} shows the evolution of the carrying capacity for different diets.
Below $\bar\beta$ the carrying capacity depends almost linearly on the diet: a diet that uses five times more soil leads to a decrease in the carrying capacity of approximately five times.
However, in the nonlinear regime the land consumption is so high, and the $N_{eq}$ so small, that a change in diet towards a more virtuous behavior does not restore sustainability.
The slope of $N_{eq}(\beta,Y_0)$ is greater when $R$ is small, and smaller for diets consuming more soil.
The diets consuming more ground reduce the possibility of population growth, $N_{eq}$ is small, and an increment of $\beta$ does not change its value too much.
In some sense we could say that vegetarians are more sensitive to the urbanization process than non vegetarians, because the latter live in a regime where the carrying capacity is very small anyway.
The point where the derivative has a minimum that corresponds to the inflection point (dotted curve of Fig. \ref{fig:derivative}, and the point where the second derivative has a minimum corresponds to the inflection of the derivative and can be considered a bound for the linear behavior of the system.
For a diet consuming soil (with more meat) the carrying capacity is very low and represents a world of few people using a lot of soil to live and to eat, but does not look coherent with the demographic trend of the planet.
If $R$ is small, the environment is much more sensitive to the country depletion because it expects to be able to accommodate a larger population.

It has been noticed that ``between 1985 and 2000, the population of Accra, the capital of Ghana, increased from 1.8 to 2.7 million, a 50\% increase. Its urban land cover increased from 13,000 to 33,000 ha, a 153\% increase: urban land cover in Accra grew more than twice as fast as its population.'' \cite{citeulike:9394452}.
Although the consequences of this land use behavior are not evident from the initial part of the solution $N(t)$, when the population approaches the carrying capacity, even small fluctuations of $C$, due for example to meteorological phenomena, can cause disastrous consequences.
Moreover, the process of urbanization is essentially irreversible on historical time scales, because the production of fertile soil is a very slow process \cite{JSFA:JSFA2740141201}.

\section{The power law for world cities}
\label{powlaw}
Equation (\ref{chini}) is based on the assumption that the dependence of urban size on population is given by (\ref{pow}).
To verify this hypothesis, we analyzed the {\sl Atlas of urban expansion} \cite{angel2012atlas} where the data of 3646 cities with more than 100,000 inhabitants have been collected for the year 2002.
In Table \ref{tabella} the values of $Y_0$, $\beta$ are reported for twelve world regions together with the adjusted square $R_j^2$ and the number of observations.
We split the Europe and Japan data because Tokyo is the most populated metropolitan area in the world.

The values of $Y_0$ are also relevant, as can be easily seen observing that the population density is given by $N(t)^{1-\beta}/Y_0$: when $Y_0$ is very small, the value of $\beta$ can be very close to 1 or even slightly larger. 

\begin{table}
	\caption{The values of $Y_0$ and $\beta$ computed from the data of the {\sl Atlas of urban expansion} \cite{angel2012atlas}}
\vskip 0.5cm
\begin{tabular}{l*{5}{c}r}
Region & $Y_0$ & $\beta$ & $R_j^2$ & Observations \\
\hline
USA            	& 0.361 & 0.856 & 0.95 & 243   	\\
Europe         	& 0.071 & 0.914 & 0.87 & 693   	\\
Japan			& 0.348 & 0.809 & 0.97 & 103   	\\
Western Asia 	& 0.059 & 0.896 & 0.89 & 157	\\
South and Central Asia 	& 0.678 & 0.700 & 0.63 & 539 	\\
Southeast Asia 	& 0.011 & 0.997 & 0.84 & 196	\\
Eastern Asia and Pacific & 0.005 & 1.035 & 0.55 & 891	\\
Northern Africa & 0.111 & 0.828 & 0.81 & 115 	\\
Sub-Saharan Africa & 0.004 & 1.047 & 0.66 & 258	\\
Latin America and the Carib & 0.278 & 0.799 & 0.83 & 403	\\
Canada 			& 0.027 & 0.995 & 0.96 & 29		\\
Australia and New Zealand & 0.530 & 0.824 & 0.95 & 19\\ 
\label{tabella}
\end{tabular}
\end{table}

\begin{figure}[ht]
\centering
\includegraphics[width=1\textwidth]{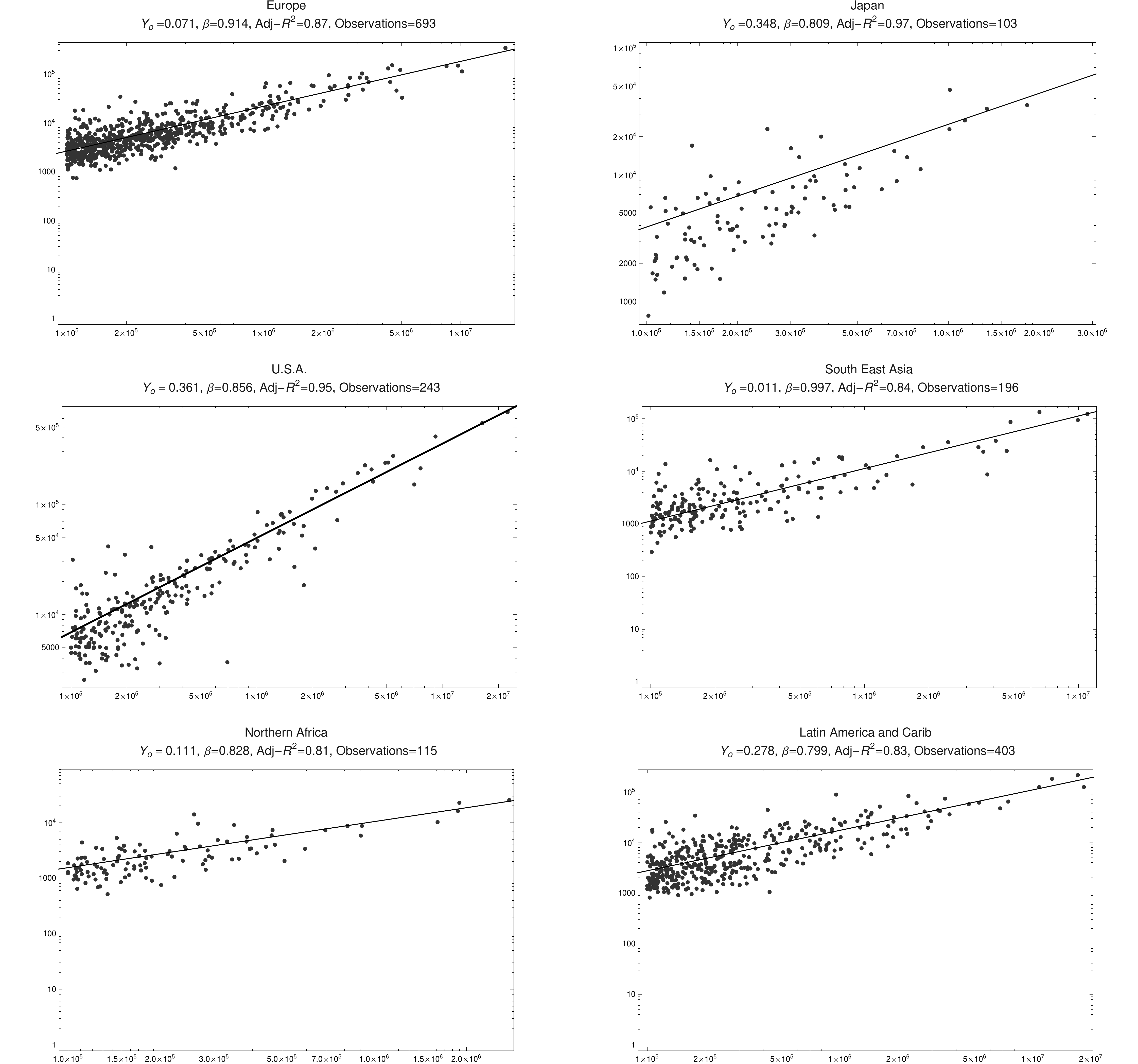}
\caption{The values of $Y_0, \beta$ for different world regions}
\label{fig:ilmondo}
\end{figure}

In Fig.\ref{fig:ilmondo} the population/urban-soil data are plotted in double logarithmic scale, and the power law (\ref{pow}) is verified with a good statistical approximation.
Even excluding the data with $R_j^2=0.55, R_j^2=0.63$ and $R_j^2=0.66$ for the Eastern Asia and the Pacific, South and Central Asia and the Sub Saharian Africa respectively, the other nine regions have an adjusted square value ranging from $R_j^2=0.81$ for Northern Africa to $R_j^2=0.99$ for Japan. 
The $\beta$ values ranges from $0.70$ for the South and Central Asia to $1.04$ for Sub Saharan Africa, and $Y_0$ from $Y_0=0.004$ for the Sub Saharan Africa ($\beta=1.04$) to $Y_0=0.67$ ($\beta=0.7$) for South and Central Asia.

To find the curve in $(Y_0,\beta)$ below which the carrying capacity is a quasilinear function of $\beta$, we computed numerically the derivatives ${\partial\over\beta} N_{eq}(\beta,Y_0)$ and ${\partial^2\over\beta^2} N_{eq}(\beta,Y_0)$, plotted in Fig. \ref{fig:derivative} without units.
If we assume that the values of $\bar\beta$ for which the second derivative has the first minimum corresponds to the curve in the plane $(Y_0,\beta)$ above which there is no sustainability,
we have a bound for the quasi linear the regime.
In this sense we say that for given $Y_0$ the urban sustainability is given by $\beta<\bar\beta$.
 
\begin{figure}[ht]
\centering
\includegraphics[width=0.8\textwidth]{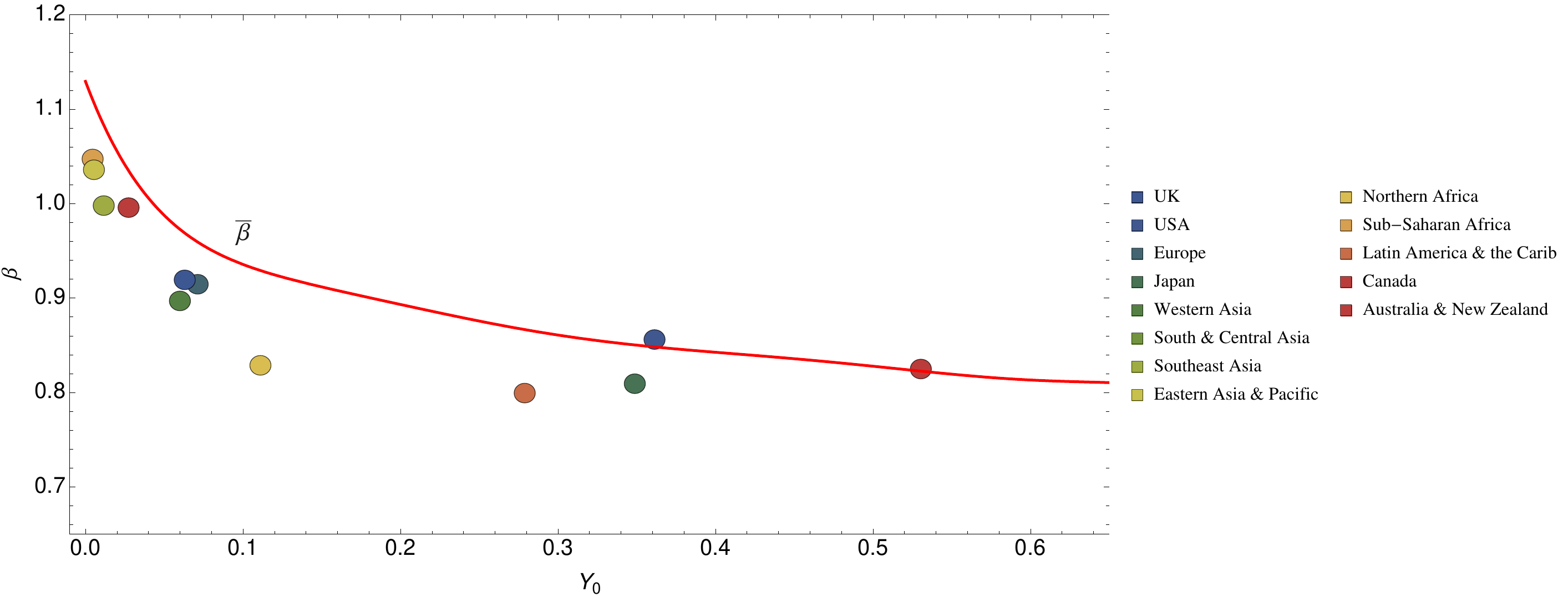}
\caption{The values $(Y_0,\beta)$ for the world regions with $R=0.18$. The red line gives the critical $\bar\beta$ for different $Y_0$ when $C=200,000$ hectares.}
\label{fig:ybeta}
\end{figure}

The shape of the curve $(Y_0,\beta)$ depends also on $C$ and $R$.
In particular if $C$ is big enough the curve is higher than the axis $Y_0$.
In Fig. (\ref{fig:ybeta}) the $(Y_0,\beta)$ values are plotted for the world regions together with the critical $\bar\beta$ as a function of $Y_0$ where the diet is given by $R=0.18$ and $C=200000$ hectars.

The city areas of our anaylis are derived from the medium-resolution Landsat satellite images on which the {\sl Atlas of urban expansion} is based \cite{angel2012atlas}.
However there is no common definition of metropolitan areas.
In \cite{10.1257/aer.101.5.2205}, for example, the city area is found by means of the City Clustering Algorithm (CCA) introduced in \cite{7032427}.
The CCA method allows one to define the city also where houses are spread in an agricultural area (urban sprawl).
The power law (\ref{pow}) relating the urban area with the population size was verified also in \cite{10.1257/aer.101.5.2205}, where the value of 1.065 was found for Great Britain and 0.958 for United States, values very close tho the $\beta=0.914$ and $\beta=0.856$ we found for Europe and the U.S.A. respectively.
In \cite{10.1257/aer.101.5.2205} it is also noted that, for opportune values of the parameter, there is a good agreement between the CCA values and those of Metropolitan Statistical Areas.
The CCA city definition includes some {\sl non urban} areas inside the metropolitan areas suggesting that the method could be useful to study the agricultural areas inside cities.
In a successive paper, we expand our model on the basis of the CCA model to evaluate the impact of urban agriculture.

\section{Conclusions and open problems}
\label{sviluppi}
The model we introduced is a simple, but nonlinear representation of the effects of urban land use on food availability.
The model depends on the hypothesis that $Y=Y_0N^\beta$, confirmed with good approximation by the experimental data.
The general form for the equation (\ref{chini}) is 
\begin{equation}
	f(t){dN\over dt}(t) = g(t)N(t)+h(t)N(t)^\beta+c(t)
\end{equation}
and the solving techniques for this kind of equation are based on Lie transform \cite{Cheb-Terrab:math-ph0007023} \cite{kamke1956differentialgleichungen}.

In general $R(t)$ is a random variable, and its behavior will be determined by its probability distribution $\rho(R)$, depending on eating habits, the latitude where the food is produced and other parameters.
The equation (\ref{chini}) can therefore be generalized to the stochastic case, and this will be the subject of a subsequent work.
The stochastic generalization of (\ref{chini}) should give a probabilistic description of the urban and rural global land use.

The functional form of $R(t)$ in eq. (\ref{chini}) also takes changes in agricultural technologies into account.
A smaller value for $R(t)$ means that the same amount of agricultural land gives food to more people, by improving technologies such as vertical farming.

The function $C(t)$ is also relevant for more detailed analysis of the sustainability.
Indeed, we have assumed that $C$ does not depend on time, but in general $C(t)$ may change over time due to weather events, loss of fertility, desertification, pollution of the soil, or phenomena of land grabbing.
We numerically solved Eq. (\ref{chini}) when $C(t)$ is a decreasing function (linear or nonlinear) and we saw that the limit of $N(t)$ for $t\rightarrow\infty$ is always dominated by the values of the function $C(t)\over R$.

Urban planning in the coming years will have to take into account the proximity of agriculture far more than has been done so far.
Strategies to strengthen the resilience of the city in fact depend also on the connecting infrastructure and transportation of agricultural products \cite{billen:hal-00645733}.
Urban growth with a large value of $\beta$ can be can be risky, as it strengthens the economic dependence of the city on the energy resources needed to carry the food.
Strategies of densification, shrinking cities or urban farming, go in the direction of reducing the value of $\beta$, in the framework of the model we proposed.
When the fossil-fuel based agriculture is no longer profitable, one needs to produce more local food if possible in terms of the carrying capacity we computed.

We considered only big cities, although, for example, in Italy, many agricultural areas are dotted with barns and houses.
In a further work we analyze the urban sprawl phenomena for the case study of northern Italy.
The data of the Region of Lombardy by municipality, for five different land uses (urban, agricultural, forest, wetland and watershed) show that Eq. \ref{pow} has been confirmed only for urban land.
In addition, we observe that an increase in population also produces a depletion of the soil (landfills, pollution, production of biofuel) independent of its agricultural use, which makes $C(t)$ a decreasing function.
The analysis of the ``land grabbing'' behavior on sustainability will be the subject of a further study.

\section{Acknowledgments}
\label{akno}
The authors would like to thank Laura Tedeschini Lalli and Paola Magrone for their useful observations, Robert Israel for his information about the Chini equation, Yota Nicolarea of FAO for bringing to our attention the phenomenon of land grabbing, Paul Blanchard and Valerio Talamanca for careful reading of this manuscript.
The statistical analysis and the numerical integrations have been done by  Wolfram Mathematica 9.1

\bibliographystyle{plain}
\bibliography{/Users/dautilia/Desktop/DTC/WorkInProgress/References/bibliografia_rob.bib} 

\end{document}